\documentstyle[12pt]{article}
\newcommand{\dba}{\not{\!{\!D}}}
\begin{document}
\title{ ON THE WESS-ZUMINO-WITTEN ANOMALOUS FUNCTIONAL AT FINITE TEMPERATURE}
\author{R.F.Alvarez-Estrada, A. Dobado and A.G\'omez Nicola
\\Departamento de F\'{\i}sica
Te\'orica, Universidad Complutense
\\28040, Madrid, Spain}
\maketitle
\begin{abstract}
We discuss the finite temperature extension of the anomalous
Wess-Zumino-Witten lagrangian. The finite temperature
$S^1\times S^3$ compactification makes a structure in disconnected
sectors, corresponding to different baryon numbers appear naturally.
The consistency of the anomalous functional
is proved for arbitrary baryon number configurations.
The anomalous behavior of the functional is shown to
be consistent with the absence of finite temperature
corrections to chiral anomalies in QCD,  for each
baryon number sector. \hspace{3cm}{\small FT/UCM/9/93}
\end{abstract}
\newpage

The low-energy properties of QCD have been successfully
studied in the
framework of the effective lagrangian method \cite{wein}\cite{ga-le},
 also called Chiral Perturbation Theory ($\chi $ PT).
In this formalism, the anomalous sector of the theory at zero temperature (ZT)
is described
through the inclusion of the Wess-Zumino-Witten (WZW) lagrangian
\cite{wz}\cite{witt} (see also \cite{ag-g} for a complete and detailed
exposition), which reproduces the correct QCD anomalies
and, consequently, the low-energy limit of anomalous processes
involving hadrons and photons, such as
$K^+K^- \rightarrow \pi^+ \pi^- \pi^0$ or $\pi^0\rightarrow\gamma\gamma$.
 The ZT WZW action cannot be expressed as a local four-dimensional integral,
but
 it has to be performed on a five-dimensional space whose boundary is the
standard
 four-dimensional space-time.
 $\chi $ PT  can  also be studied in the finite temperature
 ($T\neq 0$) case , corresponding to a hadronic medium in
 thermal equilibrium at temperature $T$. Thermodynamical quantities, like
 the pressure and the behavior of the quark condensate, have been
 analyzed in great detail in \cite{ge-le} for
 low temperatures (yet excluding anomalous WZW effects).
Analogous studies with effective models have been carried out in
\cite{gatto}. In addition to that, one may well want to study
 the  anomalous processes listed above, in the presence of the medium.
Then, a suitable
 extension of the WZW anomalous lagrangian to the $T \neq 0$ case is needed.
 In the $T=0$ case, one can compactify the euclidean space-time to the
four dimensional
 sphere $S^4$. Then, the Goldstone fields for the $N_f=3$ case ($N_f$ being
 the
 number of flavours) are mappings $U(x): S^4 \rightarrow SU(3)$
($x$ being in $S^4$). The
  lagrangian to the lowest order in the number of derivatives,
 and without any gauge coupling, is then given by the
 non-linear sigma-model one plus the WZW term \cite{witt}\cite{ag-g}:
 \begin{equation}
  \Gamma [U] = \frac{1}{4}F^2\int_{S^4} dx  Tr \{ \partial_{\mu}U^{-1}
  \partial^{\mu}
  U - M^2(U+U^{-1}) \} + \frac{N_c}{240\pi^2}\int_{I\times S^4}
  Tr(U_t^{-1}(d+d_t)U_t)^5
  \label{eq.1}
  \end{equation}

Here, $F\simeq 93 MeV$ is the pion decay constant, $M$ is the mass matrix,
$N_c$
is
the number of colours,
$I$ is the interval  $[0,1]$ in which $t$ varies ,
$d\equiv \partial_{\mu} dx^{\mu}$,
$d_t\equiv \partial_t dt$. The   last term is antisymmetrized in all the
indices, according to  the notations in \cite{ag-g}, which we shall use
in what follows almost everywhere.
 $U_t$ is a continuous mapping $U_t(x): I\times S^4 \rightarrow SU(3)$ such
 that $U_0(x)=1$, $U_1(x)=U(x)$ i.e. it is a homotopy. Since the fourth group
of
 homotopy $\pi_4 (SU(3))=0$, all the mappings $U(x)$ are deformable one
into another, and such an homotopy $U_t(x)$ always exists for any $U(x)$.
The WZW
term appearing in eq.(\ref{eq.1}) is independent of the homotopy $U_t(x)$
and, so, it is
  in fact a function of the chiral field $U(x)$ only.

In the
$T\neq 0$  case, the situation is quite different : in the
 imaginary-time formalism for the partition function \cite{lan-va},
 the euclidean time coordinate $\tau$ runs from $0$ to $\beta=1/T$ and the
fields are
 periodic in $\tau$ ($U(\vec{x},\tau+\beta)=U(\vec{x},\tau)$). Then,
 if we compactify the euclidean three-space to $S^3$ by imposing boundary
 conditions at  infinity, we deal with applications
 $U(\vec{x},\tau): S^1 \times S^3 \rightarrow SU(3)$. $\vec{x}$ will
always denote a vector in $S^3$. These applications are
 not all homotopically equivalent, i.e. they are not deformable into one
another
 as in the previous case of the $S^4$ compactification at $T=0$.
 In fact, there are
as many homotopy classes of those mappings as integer numbers \cite{white}.
The
 integer label that classifies a given map can be written as:
 \begin{equation}
 N_B=\frac{1}{24\pi^2}\int_{S^3} (U^{-1}(\vec{x},\tau)\vec{d}U(\vec{x},\tau))^3
 \label{eq.2}
 \end{equation}

 where $\vec{d}\equiv \vec{\partial}d\vec{x}$ in three dimensions and
(\ref{eq.2}) is
 antisymmetric in all the spatial indices. Notice that $N_B$ in
 (\ref{eq.2}) for
 a given $\tau$ is the winding number in $\pi_3(SU(3))$. As
 $\pi_3(SU(3))$=${\cal Z}$ (the set of all integers) also, $N_B$ cannot depend
on $\tau$,
 and it in fact characterizes topologically different maps in
 $S^1 \times S^3$. This integer $N_B$ has a natural physical interpretation
 as the
baryon number of the configuration.

 In this work, we shall consider the case in which weak gauge couplings are
neglected.
 Then, baryon number is strictly conserved. Under these conditions,
 there exist different disconnected sectors corresponding to
 $N_B=0,\pm 1,\pm 2,...$. Transitions between them are forbidden, and
 we must define our WZW lagrangian in each
 sector separately, because there is not any homotopy transformation
 connecting every $U(\vec{x},\tau)$ field with
 the identity. If gauge
 couplings with $SU(2)_L$ gauge fields were considered, the baryon number
 current would become anomalous \cite{thof} and transitions from one sector to
another
 would in principle be allowed.

  At this point, we would like to stress
that the $S^1\times S^3$ compactification scheme is useful not only to study
 finite temperature effects, but also at ZT in cases where
 it is important to have a clear realization
  of the baryon number \cite{witt}.

Let us consider now the following $S^1\times S^3$ WZW functional at finite
temperature for the $j$-th (or $N_B=j$)
sector:
\begin{equation}
  \Gamma_j[U(\vec{x},\tau);\bar{U}_j (\vec{x},\tau)]=N\int_{I\times S^1 \times
S^3}
  Tr(U_t^{-1}(\vec{x},\tau)(d+d_t)U_t (\vec{x},\tau))^5
  \label{eq.3}
  \end{equation}

$U(\vec{x},\tau)$ being in the $j$-th sector. $\bar{U}_j(\vec{x},\tau)$
is a given field also belonging to
the $j$-th sector and  hence, in the general case, it
is different from the identity, except for  $j= 0$
where it can be taken to be unity ($\bar{U}_0 (\vec{x},\tau)=1$).
$U_t$ ($t \in I$) is now a continuous application $U_t (\vec{x},\tau): I\times
S^1 \times
S^3\rightarrow SU(3)$ connecting the chiral field   $U (\vec{x},\tau)=
U_1(\vec{x},\tau)$
and the homotopy class representative, $\bar{U}_j(\vec{x},\tau)=U_0(
\vec{x},\tau)$.
$N$ is a constant to be determined.
 As in the $T=0$ case, in order to show that $\Gamma_j$ is the
 correct anomalous
 functional, it is necessary to demonstrate  that it is independent of
 the choice of the homotopy operator $U_t (\vec{x},\tau)$, i.e.,
 it depends only of the values at the boundary, and
  that it reproduces the QCD anomalies (for instance,
 the $U_A(1)$
 anomaly that gives rise to the  $\pi^0\rightarrow\gamma\gamma$ reaction).

Furthermore, in this case, it is necessary to prove also the independence of
the results on the choice of the class representative
$\bar{U}_j(\vec{x},\tau)$.

In order to show the independence of the functional
in eq.(\ref{eq.3}) on the homotopy connecting the chiral field with
the homotopy class representative, we consider the
difference $\Gamma_j [U_t ']-\Gamma_j [U_t]$.  $U_t '$ and
$U_t$ are two different mappings satisfying the required  boundary conditions,
 say, at $t=0,1$ they both are
equal to $\bar{U}_j(\vec{x},\tau)$, $U(\vec{x},\tau)$ respectively.
$\Gamma_j[U_t]$ is just the right-hand-side (r.h.s.) of eq.(\ref{eq.3}) and so
on
for $\Gamma_j[U_t']$. By using
the transformation formula \cite{ag-g}:
\begin{equation}
Tr \{ (gh)^{-1}(d+d_t )(gh) \}^5=Tr (g^{-1}(d+d_t )g)^5 +
Tr(h^{-1}(d+d_t )h)^5 + (d+d_t)\alpha (g,h)
\label{eq.4}
\end{equation}
where $g$ and $h$ are applications from a five-dimensional manifold into
 the $SU(3)$
group and  $\alpha(g,h)$ is a 4-form whose detailed form can also be found
 in
\cite{ag-g},  it is possible to write:
\begin{equation}
\Gamma_j[U_t ']-\Gamma_j[U_t]=N \int_{S^1\times S^1 \times S^3}
Tr(h_t^{-1} (\vec{x},\tau)(d+d_t )h_t (\vec{x},\tau))^5
\label{eq.5}
\end{equation}

with $h_t (\vec{x},\tau)= U_t'(\vec{x},\tau)U_t^{-1} (\vec{x},\tau)$.
The $\alpha$ term vanishes due to the property $\alpha(g,1)=\alpha(1,g)=0$
for any $g$. $h_t$ is periodic in $t$, and so the above
difference is equivalent to compactify
the $I$ interval into $S^1$ (by identifying the extreme points).

If we want our quantum partition function to be insensitive to the above
variation,
 the difference in (\ref{eq.5}) must be $2\pi in$,  $n$ being an integer.
 This is
 almost trivial in the $T=0$ case, because the integral in (\ref{eq.5})
 is over $S^1 \times S^4$ and mappings of this space into $SU(3)$ are
 classified by the homotopy group $\pi_5(SU(3))$= ${\cal Z}$ \cite{white}.
 But (\ref{eq.5}), when integrated over $S^1 \times S^4$,  is just
 the expression for the winding number in $\pi_5(SU(3))$ and so,
 it is equal to
 $2\pi in$ if the proper choice of the  constant $N$ is made.
 However, in our present case things are different.
 The mappings from $S^1\times S^1 \times S^3$ into $SU(3)$ are not classified
 by $\pi_5(SU(3))$ but there are ${\cal Z}\times {\cal Z}$ classes
 \cite{white}. Hence, in principle, it is not clear whether (\ref{eq.5})
 should
 be $2\pi n i$.

 To prove that this is  really the case, we make use of the Atiyah-Singer
  index theorem  \cite{ati}
 in six dimensions, which reads \cite{ag-g} :
 \begin{equation}
\mbox{ind}\, i\!\dba[A]=\frac{-i}{48\pi^3}\int_{M_6} Tr F^3
\label{eq.6}
\end{equation}

Here, $i\!\dba$ is the usual Dirac operator for fermions coupled to a $SU(3)$
gauge field $A$ with strength field $F=dA+A^2$, and $M_6$
 is any six-dimensional
 Euclidean
 and compact manifold. ind $i\!\dba$ is the index of the Dirac operator and it
is,
 of course, an integer.
 Let us consider the space $M_6=S^1\times S^1\times S^4$,
 which can be decomposed
 as the union of eight patches isomorphic to ${\cal R}^6$, namely
 $I^{\pm}\times I^{\pm}\times D_4^{\pm}$, where
 $D_4$ is a four-dimensional disk and $I$ is the unit interval.
 A given gauge field is only well defined
  over each patch (for example $I^{+}\times I^{+}\times D_4^{-}$). Two gauge
fields $A_1$, $A_2$, corresponding to two
  different patches, must differ by a gauge transformation defined on the
intersection
  of the patches i.e. $A_2=g^{-1}(A_1+d)g$, $g$ being an application from the
  intersection of the corresponding patches  into $SU(3)$. Let us consider the
following gauge field
  configuration : $A^{\pm\pm +}=g^{-1}dg$ ; $A^{\pm\pm -}=0$ where $g$ is
  an arbitrary map $g(x):S^1\times S^1\times S^3\rightarrow SU(3)$.
   The superscripts in the $A$'s fields denote the patches in
   which those fields are defined.
 It can be checked that this is a well defined gauge field, since
the appropiate consistency relations are fulfilled by the transition functions
on all  intersections of three or more patches.
  By definition of the Chern-Simons form $Q_5 [A(x),F(x)]$ \cite{ag-g},
   $x$ being now in $M_6$, we have
  over any single  patch $Tr F^3=dQ_5 [A,F]$ (note that this is not
true over the whole manifold). Furthermore, $Q_5[0,0]=0$ and
$Q_5[g^{-1}dg,0]=\frac{1}{10}\int_{S^1\times S^1\times S^3}
Tr(g^{-1}dg)^5$. Then, by writing the r.h.s. of (\ref{eq.6})
  for the above gauge field configuration as
  a sum over the different patches and, then, applying the Stokes theorem in
  each patch, we arrive to the desired result, namely:
\begin{eqnarray}
\frac{\Gamma_j[U_t ']-\Gamma_j[U_t]}{2\pi i}\in  {\cal Z}& \mbox{if}
& N=\frac{m}{240\pi^2}
\label{eq.7}
\end{eqnarray}

$m$ being an integer. Another way to prove this
is to take $M_6=S^2\times S^4=\sum D_2^{\pm}
\times D_4^{\pm}$ and the gauge field configuration $A^{\pm +}=g^{-1}dg$ ;
$A^{\pm -}=0$ with $g(x):S^2\times S^3\rightarrow SU(3)$, with the restriction
that $g$ takes the same value in the two poles of $S^2$. Then, it is equivalent
to an arbitrary map of $S^1\times S^1\times S^3$ into $SU(3)$. Notice that
the
result (\ref{eq.7}) is independent of the choice of class
representative $\bar{U}_j(\vec{x},\tau)$.

As discussed above, the next thing we have to study is the transformation
 properties of the functional in eq.(\ref{eq.3}). In order to do that,
we consider two arbitrary representatives of the $j$-th class
 $\bar U^A_j(\vec x,\tau)$ and
$\bar U^B_j(\vec x,\tau)$. For any chiral field $U(\vec x,\tau)$ belonging
to the $j$-th class  there exist homotopies  $U^A_t(\vec x,\tau)$ and
 $U^B_t(\vec x,\tau)$ interpolating between the chiral field and
 $\bar U^A_j(\vec x,\tau)$ and
$\bar U^B_j(\vec x,\tau)$ respectively, so that $U^A_0=\bar U^A_j, U^A_1=U,
U^B_0=\bar U^B_j $ and $  U^B_1=U$. It is also clear that
$\bar U^{A-1}_j\bar U^B_j$ belongs to the trivial class or, in other words,
there
is one homotopy $\omega_t$ connecting $\bar U^{A-1}_j\bar U^B_j$ with the
identity and, then, we have $\omega _0=\bar U^{A-1}_j\bar U^B_j $ and
$ \omega_1=1$.

Now we consider
the expression (\ref{eq.4}) for the  change of the
Chern-Simons form for a pure gauge field under a gauge
transformation, and we
apply this formula to the particular case where $g=U_t^A$
and $h=\omega_t$, together with the definition of our functional
in eq.(\ref{eq.3}) $\Gamma_j [U_t]=\Gamma_j [U;\bar U_j]$. We find:
\begin{equation}
\Gamma_j[U^A_t\omega_t]=\Gamma_j[U^A_t]+\Gamma_j[\omega_t]+\int_{I\times S^1
\times S^3}(d+d_t)\alpha(U^A_t,\omega_t)
\end{equation}
As it was discussed above, the functional $\Gamma_j [U_t]$ depends only
on the values that $U_t$ takes at $t=0$ and $t=1$ and, in addition,
$U^A_t\omega _t$ is equal to  $U^B_t$ at $t=0$ and $t=1$. Then, we can
write:
\begin{equation}
\Gamma_j[U^A_t\omega _t]=\Gamma_j[U^B_t]=\Gamma_j[U;\bar U^B_j]
\end{equation}
We also have:
\begin{equation}
\Gamma_j[U^A_t]=\Gamma_j[U;\bar U^A_j]  ; \quad
\Gamma_j[\omega _t]=-\Gamma_j[\bar U^{A-1}_j\bar U^B_j;1]
\end{equation}
So, we can write:
\begin{equation}
\Gamma_j[U;\bar U^B_j]=\Gamma_j[U;\bar U^A_j]-\Gamma_j[\bar U^{A-1}_j
\bar U^B_j;1]
-\alpha(\bar {U}_j^A,
\bar{U}_j ^{A-1}d \bar{U}^B_j)
\end{equation}
where $\alpha(g,1)=0$ has again been used. Therefore,
we arrive to the conclusion
that $\Gamma_j[U;\bar U ^B_j]-\Gamma_j[U,\bar U^A_j]$ does not depend
on $U$. Thus, if we
introduce another chiral field $U'$ belonging also to the $j$-th class, we get:
\begin{equation}
\Gamma_j[U';\bar U ^B_j]-\Gamma_j[U;\bar U_j ^B]=\Gamma_j[U';\bar U_j^A]-
\Gamma_j[U;\bar U_j^A]
\label{eq.12}
\end{equation}
This is a very interesting result, since it shows that the variations
of the functional in eq.(\ref{eq.3}) corresponding to small variations
(in the topological sense) of the chiral field $U$ do not depend at all
on the class representative $\bar U_j$. This applies of course also
to the anomalous variations or the Green functions obtained
from this functional. The independence of Green functions on $\bar{U_j}$
 can be seen, for instance, by expanding
$\Gamma_j [U;\bar{U_j}]$ near a classical field
 configuration $U=U_c$ in order to get
the effective action including loops, and by taking into account
eq.(\ref{eq.12}).
 As a particular case, we can consider the trivial sector $j=0$ and  choose
the identity as the class representative. Now, we can write the chiral field
$U$ in terms of the Goldstone boson fields $\pi_a$ in the standard way as
$U=\exp (i\sum _a\lambda_a\pi_a/F)$. Then it is possible to make a
derivative expansion in
eq.(\ref{eq.3}) retaining only the lowest (four) derivatives. Finally,
by using the Stokes theorem, we get an action which is the
integral
on
$S^1\times S^3$ of a lagrangian which properly reproduces
processes
like $K^+K^- \rightarrow \pi^+ \pi^- \pi^0$.

Now we want to see that the functional in eq.(\ref{eq.3}) reproduces
the QCD anomalies and, in particular, the $U_A(1)$ anomaly
responsible for the $\pi^0\rightarrow\gamma\gamma$ decay. In order to
show that, let us  gauge $\Gamma_j[U;\bar{U_j}]$ with the electromagnetic field
$A_{\mu}$,
 to make it invariant under infinitesimal local charge transformations
 $U(\vec{x},\tau)\rightarrow U(\vec{x},\tau)+i\varepsilon(\vec{x},\tau)
 [Q,U(\vec{x},\tau)]\equiv U'$ with $Q$ the quark charge matrix. We start
 from $\Gamma_j[U';\bar{U}_j]$ written in (\ref{eq.3}) in terms of $U'_t$
 such that $U'_0=\bar{U}_j(\vec{x},\tau)$ and $U'_1=U'(\vec{x},\tau)$. But
 $\Gamma_j[U'_t]=\Gamma_j[U_t\omega_t]$ with $U_0=\bar{U_j}$ ,
 $U_1=U$, $\omega_0=1$ ,
 $\omega_1=1+v$ and $v=i\varepsilon(U^{-1}QU-Q)$ and so, by applying again
 (\ref{eq.4}) we obtain :
 \begin{equation}
 \Gamma_j [U';\bar{U}_j ]-\Gamma_j [U;\bar{U}_j]=\Gamma_j[1+v;1]+\alpha(U,1+v)
 \label{eq.8}
 \end{equation}
where the $\alpha$ term is four-dimensional and it is the $t=1$ contribution.

Again the $t=0$ one vanishes ($\alpha(g,1)=0$). As we have seen in the general
case in eq.(\ref{eq.12}),
 the variation of $\Gamma_j$ has  always
the same form for each sector, independently of $\bar{U}_j$. On the
other hand, the r.h.s of (\ref{eq.8}) when expanded in $\varepsilon(x)$
is the four-dimensional $T=0$ part \cite{witt} with the replacement
$S^4\rightarrow S^1\times S^3$. Then, the gauged WZW functional has the
same dependence on $A_{\mu}$ for every sector. The same argument applies
if we introduce an external $U_A(1)$ field $a_{\mu}$
to derive the axial current
in the usual way as $J_{\mu}^A=\frac{\delta
\Gamma_j[U,A_{\mu},a_{\mu}
;\bar{U}_j]
}{\delta a_{\mu}}$. As the dependence of
 the gauged functional $\Gamma_j[U,A_{\mu},a_{\mu}
;\bar{U}_j]$ on $a_{\mu}$ is the same for each sector, $J_{\mu}^A$ has
the same form in terms of each $U$ field, which turns out to be also the $T=0$
form
with the replacement commented above. Then, the anomaly for
$\partial^{\mu}J_{\mu}^A$ when $T\neq 0$ remains the same as in QCD
for $T=0$ in terms of the
$A_{\mu}$ field by choosing $m=N_c$, with the above replacement. Therefore, it
has been
shown that the WZW term in eq.(\ref{eq.3}) is consistent with other previous
results on QCD anomalies at finite temperature \cite{rfae}.

As a consequence of the above discussion,  we arrive at the
 conclusion that the functional, in eq.(\ref{eq.3}) is the proper
generalization
of the WZW functional when one uses the compactification $S^1\times S^3$
instead of the original  $S^4$. The $S^1\times S^3$ compactification
seems to be more appropriate to study either finite temperature effects,
or configurations with baryon number different from zero or both
simultaneously.
However, one could still ask about the role that the class representative
$\bar U_j$ plays in the definition of the functional in eq.(\ref{eq.3}). This
functional effectively depends on $\bar U_j$ but, as we have shown
in this work, the variation $\delta \Gamma_j[U;\bar U_j]$ produced by a
variation
in the chiral field $\delta U$ is independent on the class
 representative $\bar U_j$. Then, all  contributions to
the Goldstone boson Green functions coming from the functional
 in eq.(\ref{eq.3}) do not depend at all on $\bar U_j$.
Thus the particular choice of the
class representative has no  effect on  observables such as
widths or cross-sections. In some sense. the class representative is as
irrelevant
as the value of a constant added to the interaction lagrangian
of a $\lambda \Phi ^4$ theory.

 Following this analogy, and noting that $\Gamma_j [U;U]=0$, one
 could think that a natural choice for the
class representative $\bar U_j$ could be some static field configuration
 minimizing the energy, i.e., representing the vacuum of this sector or, in
 other
words, representing the state of the theory with minimal energy for fixed
baryon number $N_B=j$. For example, for $j=0$ this state would be the QCD
vacuum which corresponds to $\bar U_0=1$, thus recovering the standard
 WZW functional (but with different boundary conditions for the chiral field
and
 with the $T$-dependence in the $T\neq 0$ case).
Let us consider now the case of the non-trivial sectors. For instance,
 we can consider the $j=1$ class. Obviously, the state of minimal energy
with baryon number equal to one should describe the nucleon. However, a nucleon
can be located at  any place of space and, therefore, we do not have
 a single state
but many of them, degenerated and connected by simple spatial translations. The
description
of these  states  in terms of chiral fields is, clearly, some kind of skyrmion
configuration located in some given point of the space. However, to
obtain the  concrete form of this configuration one needs to compute
the effect of the higher derivative terms appearing in the $\chi PT$
lagrangian.
It is clear that, in practice, this can only be achieved in an approximate way,
for
 instance, by retaining only a few of the lowest derivative terms.  Once
an appropriate description of the chiral field describing the nucleon
be available, the functional in eq.(\ref{eq.3}) could be used to
compute anomalous
reactions in  the $j=1$ sector, such as the decay of pseudoscalar excitations
of the nucleon into two photons and one non-excited proton. Of course, this
discussion can be extended to the low energy states corresponding to
higher baryon numbers, which could in principle provide a description of
 other nuclei.

One may also calculate the effects of loops in $\chi$PT in the trivial sector,
by taking into account the functional of eq.(\ref{eq.3}). The compactification
$S^1\times S^3$ allows us to carry out these calculations at finite temperature
and/or finite volume (the total action for computing loops is now the obvious
finite temperature extension  of the first integral in the r.h.s. of
(\ref{eq.1}) plus (\ref{eq.3})). For instance, it is possible to obtain
 non trivial finite-$T$ corrections to the decay of the $\pi^0$ into two
photons
 \cite{dfg}. This is important in a theoretical sense, because one must check
 that there are no infinite contributions that would change the constant $N$
in (\ref{eq.7}) due to renormalization effects.  Also,
from a phenomenological point of view, the $\pi^0$ could be living in a
 hadronic medium and temperature effects affecting its decay can be analyzed.
 It is important to recall, that in the finite volume case, one must take
into account the fact that the chiral limit $M\rightarrow 0$ and the
infinite volume one $V\rightarrow\infty$, $V$ being the volume, do not conmute
 (see \cite{leut}). At finite $V$ there is no spontaneous symmetry breaking,
and
 the limit $M\rightarrow 0$ is inconsistent. The symmetry must be
explicitely broken with the mass term and then the infinite volume limit
can be taken. This is a crucial point here, because the $R\rightarrow
\infty$ limit, $R$ being the radius of $S^3$, must be taken at some stage.
 This means, of course, that our actual extension of the WZW
functional allows to study
in general the $T\neq 0$ case with infinite volume.

To conclude, we would like to summarize the main results of this work:
The following functional :

\begin{eqnarray}
\Gamma_j [U,\bar{U}_j]=\frac{N_c}{240\pi^2}\int_{0}^{1} dt\int_{S^1\times
S^3} \epsilon^{ijklm}U_t^{-1}\partial_i U_t U_t^{-1}\partial_j U_t
 U_t^{-1}\partial_k U_t U_t^{-1}\partial_l U_t U_t^{-1}\partial_m U_t
\nonumber\\
U_0=\bar{U}_j (\vec{x},\tau) \quad U_1=U(\vec{x},\tau) \quad i...m=1,...,5
\label{eq.14}
\end{eqnarray}
which is a more explicit form of (\ref{eq.3}),
provides the proper version of the WZW functional
 for $S^1\times S^3$ compactification of the space time reproducing correctly
the QCD anomalies. This compactification is appropriate for finite
temperature or/and finite volume computations, so that we can reproduce
previous
 results in
QCD anomalies and study processes
like
$\pi^0\rightarrow\gamma\gamma$, both at finite $T$ and/or finite $V$.
It also provides a clear
 description of the baryon number in $\chi PT$. In particular,
the functional in eq.(\ref{eq.14}) (or its gauged version with the
electromagnetic field)
could in principle be used to compute anomalous processes involving
states with non zero baryon number. One could also
envisage the possibility of gauging the $SU(2)_L$ group. In this case,
the baryon number becomes anomalous and it is not conserved any more, specially
at very high temperatures.
However, the discussion in this case is  more involved and it will
be presented elsewhere.

We are grateful to L.A.Ibort for some discussions and help.
The financial support
of C.I.C.Y.T (Projects AEN90-0034 and AEN93-0776), Spain, is acknowledged.

\end{document}